# Production of high purity TeO$_2$ single crystals
# for the study of neutrinoless double beta decay

C. Arnaboldi[a,b], C. Brofferio[a,b], A. Bryant[c,d], C. Bucci[e], L. Canonica[f,g], S. Capelli[a,b],
M. Carrettoni[a,b], M. Clemenza[a,b], I. Dafinei[h,*], S. Di Domizio[f,g], F. Ferroni[i,h], E. Fiorini[a,b], Z. Ge[j],
A. Giachero[a,b], L. Gironi[a,b], A. Giuliani[k,b], P. Gorla[e], E. Guardincerri[e,g], R. Kadel[c], K. Kazkaz[l],
L. Kogler[c,d], Y. Kolomensky[c,d], J. Larsen[c], M. Laubenstein[e], Y. Li[m], C. Maiano[a,b], M. Martinez[a,b], R. Maruyama[n], S. Nisi[e], C. Nones[k,b], Eric B. Norman[l,o], A. Nucciotti[a,b], F. Orio[i,h], L. Pattavina[a,b],
M. Pavan[a,b], G. Pessina[a,b], S. Pirro[a,b], E. Previtali[a,b], C. Rusconi[k,b], Nicholas D. Scielzo[l], M. Sisti[a,b],
Alan R. Smith[c], W. Tian[m], M. Vignati[i,h], H. Wang[m], Y. Zhu[j]

[a]*Dipartimento di Fisica dell'Università di Milano-Bicocca I-20126 - Italy*
[b]*Sezione INFN di Milano-Bicocca I-20126 - Italy*
[c]*Lawrence Berkeley National Laboratory, Berkeley, CA 94720 USA*
[d]*Department of Physics, University of California, Berkeley, CA 94720 - USA*
[e]*Laboratori Nazionali del Gran Sasso, I-67010, Assergi (L'Aquila), I-67010 - Italy*
[f]*Dipartimento di Fisica dell'Universita' di Genova I - 16126 - Italy*
[g]*Sezione INFN di Genova, Genova I-16146 - Italy*
[h]*Sezione INFN di Roma, P-le Aldo Moro 2, Roma I-00185 - Italy*
[i]*Università "La Sapienza", Dipartimento di Fisica, P-le Aldo Moro 2, Roma I-00185 - Italy*
[j]*Shanghai Institute of Ceramics Chinese Academy of Sciences, Jiading district, Shanghai 201800, P. R. China*
[k]*Dipartimento di Fisica e Matematica dell'Università dell'Insubria, Como I-22100 - Italy*
[l]*Lawrence Livermore National Laboratory, Livermore, CA 94550 - USA*
[m]*Shanghai Institute of Applied Physics Chinese Academy of Sciences, Shanghai 201800, P. R. China*
[n]*University of Wisconsin, Madison, WI 53706 - USA*
[o]*Department of Nuclear Engineering, University of California, Berkeley, CA 94720 USA*

*\* Corresponding author: I. Dafinei; INFN, Sezione di Roma; P-le Aldo Moro 2, 00185-Roma, ITALY*
*Tel.: +390649914333; fax: +39064454835; e-mail: [ioan.dafinei@roma1.infn.it](mailto:ioan.dafinei@roma1.infn.it)*

**Abstract**
High purity TeO$_2$ crystals are produced to be used for the search for the neutrinoless double beta decay of $^{130}$Te. Dedicated production lines for raw material synthesis, crystal growth and surface processing were built compliant with radio-purity constraints specific to rare event physics experiments. High sensitivity measurements of radio-isotope concentrations in raw materials, reactants, consumables, ancillaries and intermediary products used for TeO$_2$ crystals production are reported. Production and certification protocols are presented and resulting ready-to-use TeO$_2$ crystals are described.

______________________________________________________________

______________________________________________________________

**1. Introduction**
In recent years, much has been learned about the properties of neutrinos. From studies of atmospheric [1], solar [2], and terrestrial [3] neutrinos, we now know that neutrinos oscillate between "flavor" eigenstates, i.e. electron-type, muon-type, and tau-type. This type of oscillation can only take place if neutrinos have finite masses. However, oscillation experiments are sensitive only to differences in the squares of neutrino masses, not to the absolute scale of the neutrino mass. More important, we still do not know if neutrinos are of Dirac type (i.e. particle and antiparticle are distinct) or of Majorana type (i.e. particle = antiparticle). Studies of the rare radioactive decay process known as double beta decay (DBD) offer the possibility to address both of these outstanding issues.
In analogy to the normal beta-minus, beta-plus, and electron-capture decays, there are four different modes of double beta decay. In double beta minus decay (a process that can occur for even-even nuclei on the neutron-rich side of stability valley) two neutrons in a nucleus simultaneously convert into two protons with the emission of two electrons and, in the standard allowed process, two electron anti-neutrinos. For neutron-deficient even-even nuclei, the possibility exists for double beta-plus decay, beta-plus/electron capture decay, and double electron capture decay. In these three processes, two protons are converted into two neutrons and in the allowed versions of these decays, two electron neutrinos are emitted. If neutrinos are of Majorana type, then it is possible that all four of these double beta decay modes could occur without the emission of neutrinos. Searches for these "0ν" double beta decay processes are the subjects of much current and proposed experimental research.

If neutrinos are emitted in a double beta decay, then they share the decay energy with the emitted electrons or positrons. On the other hand, if no neutrinos are emitted, then the electrons or positrons carry away the full decay energy. Thus, the signature of 0ν double beta minus decay is a monoenergetic line in the summed electron energy spectrum at the $Q_{\beta\beta}$ value (total energy released in the decay) for the nuclide under study. Of course, the finite energy resolution of any real detector will smear this line out into a narrow peak. There are several techniques being used to search for 0ν DBD by measuring the energies of the emitted charged leptons. Conventional ionization detectors measure the charge produced in a material as the electrons or positrons come to rest. An alternative that is being used in the Cryogenic Underground Observatory for Rare Events (CUORE) is the bolometric technique, in which the heat produced by the stopping of the electrons or positrons is measured. In CUORE, the detector is also the source of the DBD events.

In order for the energy deposited by a typical radioactive decay to produce a measurable temperature rise, the heat capacity of the absorber must be extremely low. For a single crystal at low temperatures, the Debye law states that the heat capacity varies as $T^3$. Thus, large crystals of material containing double beta decaying nuclei can be used as very sensitive detectors for 0ν DBD.

## 2. TeO$_2$ use for the study of neutrinoless double beta decay

In principle, almost any crystal containing DBD candidates could be used in cryogenic experiments. TeO$_2$ crystals have the advantage of very good thermal and mechanical properties. Furthermore, the high natural abundance of $^{130}$Te (33.8%) eliminates the need for complicated and costly enrichment processes.

### 2.1. TeO$_2$ general properties

Tellurium dioxide is found in nature in two mineral forms tellurite (orthorombic) and paratellurite (α-TeO$_2$). The latter has a tetragonal symmetry $D_4(422)$ and due to its useful acousto-optic properties, is commercially produced on a large scale. First crystal synthesis using Czochralski method was reported in 1968 [4]. The first Bridgman growth of TeO$_2$ crystals was reported in 1985 [5]. The structure of the crystal may be regarded as a distorted rutile structure with a doubling of the unit cell along the [001] direction where the positions of the tellurium are slightly shifted from the regular rutile positions [6]. Tellurium ion is fourfold coordinated by oxygen, the coordination polyhedron being a distorted trigonal bipyramid [7] with two different bond distances: 1.88 Å (short) and 2.12 Å (long). Each oxygen atom is bonded to two tellurium atoms with an angle of 140° between long and short bonds; other tellurium neighbors are at larger distances. Oxygen positions have no symmetry. Eight nonequivalent oxygen or interstitial sites and four nonequivalent tellurium sites can be distinguished. The material is birefringent, optically active and highly transparent [8] in the range of 350 nm – 5 μm. TeO$_2$ is insoluble in water and has high refractive indices ($n_o$ = 2.274 and $n_e$ = 2.430 at λ = 500nm). The density of TeO$_2$ synthetic crystals is 6.04 g/cm$^3$, close to the density calculated from measured lattice constants: a = 4.8088 Å and c = 7.6038 Å. The melting point of commercially available TeO$_2$ 99.99% purity powder is 733°C.

TeO$_2$ crystal is particularly convenient for use in a cryogenic particle detector because of its thermodynamic characteristics. TeO$_2$ is a dielectric and diamagnetic crystal for which, at low temperature, the specific heat, c, can be calculated using the Debye law (c ~ $(T/T_D)^3$). The relatively high value of Debye temperature, $T_D$ = 232 K [9] yields a very low heat capacity at cryogenic temperatures leading to large temperature variations and good energy resolution. Moreover, the thermal expansion coefficient of TeO$_2$ crystal [10] is very close to that of copper, allowing copper to be used for the mechanical support structure of the detector without placing too much strain on the crystals in the cooling process.

### 2.2 Specific requirements for rare events physics applications

The crystal absorber forms the core of a cryogenic bolometer. The energy deposited by a particle is measured through the temperature increase of the crystal. In DBD experiments like CUORE, the crystal is both the source of the decay and the bulk of the detector. In principle, DBD events can be identified by their energy spectrum. However, in practice, the capability of the experiment to identify the DBD signal is limited by the presence of environmental radioactivity and radioactive contaminants in the detector. Therefore, it is critical that the detector be free of any contaminant that can mimic or simulate the DBD signal by producing an energy deposit near the DBD $Q_{\beta\beta}$ value.

The contamination may come from long-lived, naturally occurring isotopes, such as $^{238}$U, $^{232}$Th, $^{40}$K and their daughters and from cosmogenic activation of the detector materials after their production. To minimize the influence of long-lived nuclei, great care must be devoted to the selection of all materials and ancillaries used for the preparation of the detector. In the case of TeO$_2$ crystals, this means a detailed monitoring of all materials, tools and facilities used for raw oxide synthesis and crystal growth. Possible surface contamination must be controlled with great care by a strict selection of consumables and equipment used for chemical and mechanical processing of grown crystals. Materials and tools used in the packaging process must also be strictly controlled for the same reasons. Sea level transport and underground storage of prepared crystals are necessary in order to minimize their cosmogenic activation.

Based on previous experience [11] and on a background goal of ($10^{-2}$ -- $10^{-3}$) counts/keV/kg/year for the CUORE experiment [12], limit values were defined for the concentration of radionuclides to be accepted in raw materials, consumables, reagents and intermediary products used for the production of $TeO_2$ crystals for CUORE. The values are reported in Table 1.

**Table 1** Concentration limits for radioactive isotopes requested for raw materials, reagents, consumables and intermediary products used for the production of $TeO_2$ crystals

| material | category | contamination limits |
|---|---|---|
| metallic Te | raw material | $^{238}U < 2*10^{-10}$ g/g<br>$^{232}Th < 2*10^{-10}$ g/g<br>$^{210}Pb < 10^{-4}$ Bq/kg<br>$^{40}K < 10^{-3}$ Bq/kg<br>$^{60}Co < 10^{-5}$ Bq/kg |
| water and acids used for $TeO_2$ powder synthesis | reagent | $^{238}U < 2*10^{-12}$ g/g<br>$^{232}Th < 2*10^{-12}$ g/g |
| water | consumable | $^{238}U < 2*10^{-12}$ g/g<br>$^{232}Th < 2*10^{-12}$ g/g |
| $TeO_2$ powder before crystal growth | intermediary product | $^{238}U < 2*10^{-10}$ g/g<br>$^{232}Th < 2*10^{-10}$ g/g<br>$^{210}Pb < 10^{-4}$ Bq/kg<br>$^{40}K < 10^{-3}$ Bq/kg<br>$^{60}Co < 4*10^{-5}$ Bq/kg<br>Pt $< 10^{-7}$ g/g<br>Bi $< 10^{-8}$ g/g |
| $TeO_2$ crystal, ready-to-use | final product | $^{238}U < 3*10^{-13}$ g/g<br>$^{232}Th < 3*10^{-13}$ g/g<br>$^{210}Pb < 10^{-5}$ Bq/kg<br>$^{60}Co < 10^{-6}$ Bq/kg |
| crystal polishing ($SiO_2$ powder and textile pads) | consumables | $^{238}U < 4*10^{-12}$ g/g<br>$^{232}Th < 4*10^{-12}$ g/g |
| gloves, plastic bags, cleaning tissues, etc | ancillaries | $^{238}U < 4*10^{-12}$ g/g<br>$^{232}Th < 4*10^{-12}$ g/g |

## 3. $TeO_2$ production and certification procedures

$TeO_2$ crystals currently produced at industrial scale for acousto-optics applications have mechanical and optical characteristics fully compliant with DBD application requirements. In this case, crystal quality is driven by radio-purity constraints, i.e. very low level, possibly absence of radioactive nuclides including those potentially produced by cosmic rays activation. A dedicated protocol was defined in the present work for the radio-purity related quality control of the crystal production process starting from metallic tellurium synthesis to the final processing of ready-to-use $TeO_2$ crystals.

### 3.1 Radio-pure $TeO_2$ crystals production

The production process of $TeO_2$ crystals for DBD use is divided into two major phases, crystal synthesis and crystal polishing, which are then divided into several sub-processes. Bulk contamination is a risk in the crystal synthesis phase, while surface contamination is the main concern during the crystal processing.

The raw material synthesis and $TeO_2$ crystal growth methodology applied in the present work were basically described in previous works [13, 14]. The key to obtaining high purity crystals is making two successive crystal growth processes with two associated iterations of $TeO_2$ powder synthesis, as illustrated in Fig. 1. The $TeO_2$ powder used for the first growth is synthesized from metallic Tellurium dissolved with aqua regia ($HNO_3$: HCl = 1:3) and then precipitated with concentrated ammonia. After the washing and drying processes, the obtained raw dried (RD) powder is further calcinated at 680 °C in Pt crucibles for 24h in free atmosphere thus obtaining the raw calcinated (RC) powder used for the crystal growth. $TeO_2$ seeds used for the crystals growth along <110> direction are cut and oriented within 30' precision, shaped and washed using ultrapure water and reagents, following a dedicated protocol. For the crystal growth, platinum crucibles are filled with the crystal seed and the RC powder, sealed and placed into alumina refractory tubes which at their turn are placed into modified Bridgman furnaces [14]. Crucibles are heated to about 800-860 °C and kept at this temperature for several hours after which they are raised to a certain height in order to melt the top of the seed and keep the system still for 4h in order to create a stable solid–liquid interface. The growth process is then driven by lowering the crucible at a rate of 0.6 mm/h and raising the furnace temperature by about 3 °C/h. At the end of the growth process the furnaces are cooled down to room temperature slowly in order to avoid crystals' cracks caused by thermal stress. The high purity $TeO_2$ powder used for the second growth is prepared from selected regions of crystals obtained in the first growth. Selected $TeO_2$ crystalline pieces are dissolved with concentrated HCl and then precipitated with concentrated ammonia. The precipitate is further washed and dried at 80 °C. The ultra-pure dried (UPD) powder thus obtained is further treated at 680 °C in a Pt crucible for 24h in free atmosphere to obtain the ultra-pure calcinated (UPC) powder and a second growth process is executed following the same procedure as in the first one. Nevertheless the tools and ancillaries used for the two growth processes are strictly separated in the frame of a radio-pure crystals production protocol applied in the case of $TeO_2$ crystals for DBD application.

In the second phase of $TeO_2$ crystal production, the raw crystal ingots are subject to a rough mechanical processing (cutting, orienting and shaping) followed by the final surface treatment and packaging. A dedicated clean room was built for this purpose [15], and a processing protocol was implemented which includes chemical etching of crystal surfaces. A special vacuum packaging procedure was also defined in order to reduce surface radio-contamination risks, especially due to radon exposure.

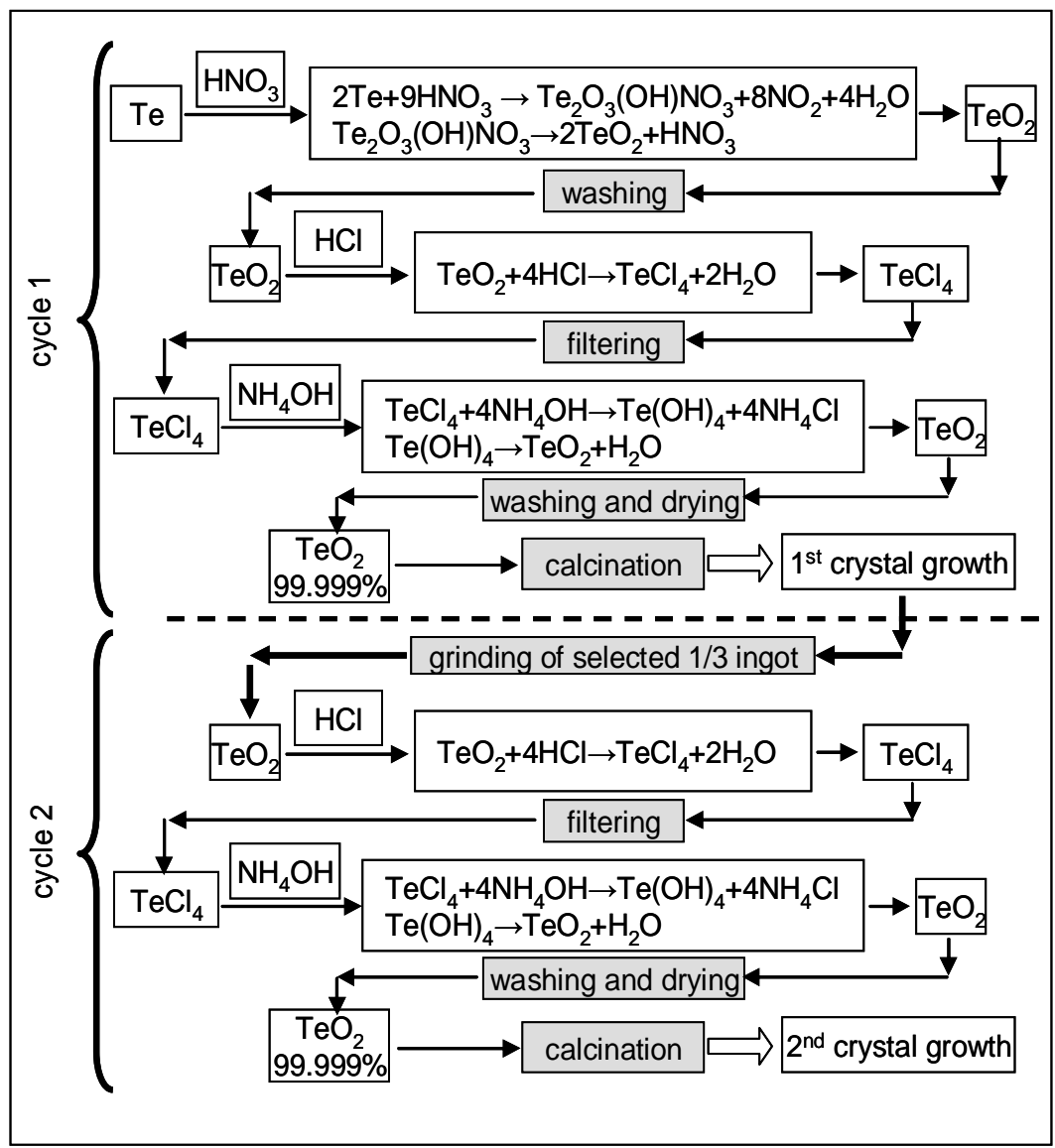

Fig. 1 Raw material and crystal synthesis protocol applied for the production of TeO$_2$ crystals.

The entire production process is subject to a complex validation protocol illustrated in Fig. 2 which includes radio-purity certification procedures to be applied in each production phase and immediate actions to be taken in case of failure, including the halt of crystal production until the problem is solved.

**3.2 Measurement methods for radio-purity certification**

Radio-purity certification included a strict control of consumables, equipment and procedures for each production phase as well as measurement of radioactive nuclide concentrations in raw materials, reagents and intermediary products during the whole process of TeO$_2$ crystal synthesis. The different measurement techniques used for radio-purity certification are briefly discussed in the following sections.

**3.2.1 Inductively Coupled Plasma Mass Spectrometry**

Inductively Coupled Plasma Mass Spectrometry (ICP-MS) measurements were performed at Laboratori Nazionali del Gran Sasso (LNGS) on all raw materials and intermediary products and the results were systematically cross-checked following the same measurement protocol on twin samples at Shanghai Institute of Applied Physics (SINAP) in Shanghai, China and Lawrence Berkeley National Laboratory (LBNL) in the USA. Dedicated sample preparation protocols (different procedures depending on the nature of samples) were developed to reach the best possible sensitivity while limiting the salt concentration in the solution to be measured. A high concentration of salts in the measurement solution can lead to build-up on the skimmer and sampling cones, causing drift and reducing the measurement sensitivity. The same protocols for sample preparation and measurement were applied in all laboratories for cross-check measurements. In the following, some experimental details are given for the ICP-MS measurements at LNGS, where a quadrupole mass spectrometer from Agilent Technologies, model 7500a (2001), was used. A dedicated tuning of the machine was made in order to have high sensitivity, stable signals and reliable results concerning matrix effects. A Babington-type nebulizer was used, which supports a high concentration of dissolved solid without any clogging. For each sample, a group of three replicates were measured using an integration time of 1s for each point and taking 3 points/mass. The measurements were performed in semi-quantitative mode using a single standard solution containing 10 ppb of Li, Y, Ce and Tl. These elements cover a broad range of atomic masses and the ICP-MS device calculates the sensitivity for other elements as a function of their response. This operating mode is fully satisfactory for the present work, where contamination levels in the samples are typically at or below the detection limit. The use of a better calibration curve would possibly improve the quality of data but eventually would cause an increase of the background, reducing the obtainable sensitivity. Nevertheless, the instrumentation response has been checked directly on the elements to be measured by adding a low concentration spike (about ten times higher than the expected detection limit) to the final measurement solution. On the other side it was fundamental to check the efficiency of samples treatment procedure measuring the recovery on spiked of K, Th and U added directly to the sample (Te metal or TeO$_2$) before the chemical etching. All tests carried out for this purpose have shown that the sample treatments efficiency was close to 100%. The reagents blank and sample treatment procedure have been subtracted for all measurements. The errors of concentration values determined in semi-quantitative mode reported in this work are less than 25-30%. This range of accuracy was confirmed also by the results of sample spike test recovery.

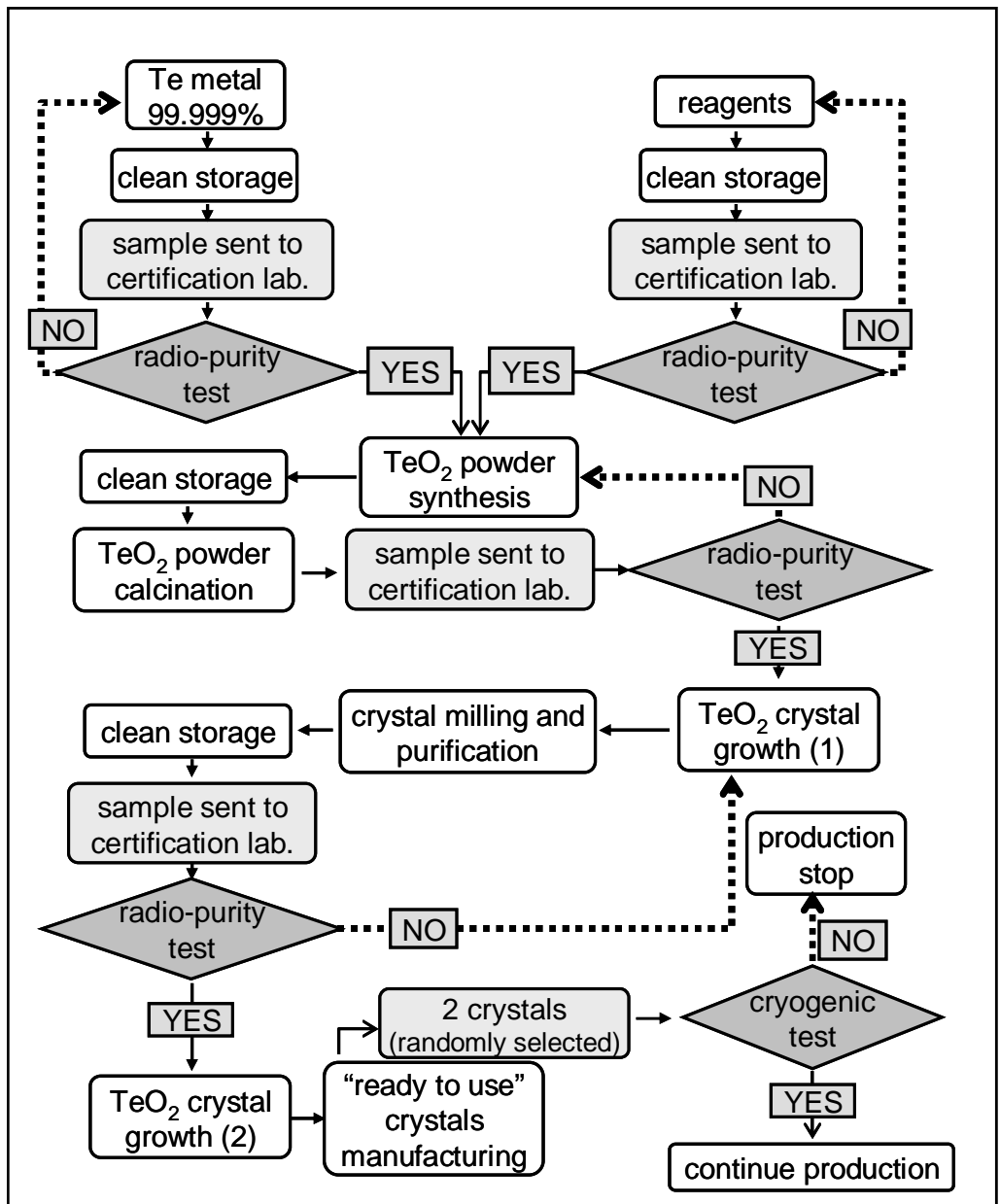

Fig. 2 Radio-purity certification protocol applied for TeO$_2$ crystals.

The sensitivities reached using this measurement protocol and dedicated samples treatment conditions were $2 \cdot 10^{-10}$ g/g for $^{232}$Th and $^{238}$U in solid raw materials and consumables (Te, $TeO_2$, $Al_2O_3$, $SiO_2$), $10^{-12}$ g/g in $H_2O$ and $10^{-11}$ g/g in all other reagents. For the other contaminants measured, the sensitivities were $10^{-9}$ g/g for $^{195}$Pt and $^{209}$Bi in solid samples and $10^{-11}$ g/g for Bi and Pb in the reagents ($10^{-12}$ g/g in the case of $H_2O$). The relatively low sensitivity ($10^{-6}$ g/g) reached by ICP-MS for $^{40}$K was compensated by complementary measurements described in Section 3.2.2.

### 3.2.2 Gamma spectroscopy with High Purity Germanium (HPGe) detectors

Gamma ray spectroscopy with HPGe detectors is one of the most important tools to determine the radioactive contamination of a sample. The concentrations of contaminants are determined by direct measurement of the gamma ray photons emitted as a consequence of their alpha and beta radioactive decay. The sensitivity of the method depends on good energy resolution for distinguishing different gamma ray lines, intrinsic low radioactive contamination of the measurement system, and effective shielding of the detector from environmental radioactivity. HPGe spectroscopy measurements reported in this work were performed at the low radioactive laboratory of the LNGS underground site [16], which offers a unique opportunity to reduce the cosmic ray background. Furthermore, the radioactive pollution in the laboratory is kept at a very low level, and specially produced radiopure HPGe detectors are used. All detectors are shielded from the environmental background with lead shields, specially selected for low levels radioactive contamination, and surrounded with special boxes, to minimize radon contamination. The measurement setup may reach sensitivities of $10^{-12}$ g/g for the U and Th chains when large samples are measured for a very long acquisition time. In our case, however, a relatively fast response time (a few weeks) is needed for timely crystal production. Previous experience showed that a lower sensitivity of $10^{-10}$ g/g for the U and Th chains was sufficient for the certification of raw materials and consumables used to obtain high radio-purity $TeO_2$ crystals, compatible with the background specifications of the CUORE experiment. 2 kg samples were typically needed for real-time measurements of the order of $2 \cdot 10^6$ sec. to obtain sensitivities on the order of $10^{-10}$ g/g for the U and Th chains. Monte Carlo simulations were performed in order to obtain the measurement efficiency. Data reported in Section 4.2 are calculated at 90% CL.

### 3.2.3 Alpha Spectroscopy with Surface Barrier Detectors (SBD)

The SBD method, able to reach a very low level counting in the alpha decay energy range, was mainly used for certification activities aimed at preventing surface contamination in the final phase of crystal production (cutting and shaping, surface processing, packaging and shipment). The study of all potential sources of surface contamination was very important in order to select the materials for surface treatments and crystal packaging. In particular, the lapping cloths and plastic bags for packaging were carefully selected to avoid all possible contaminants. Alpha decay is a good indicator of possible surface contamination from the $^{238}$U and $^{232}$Th chains. Especially in the $^{232}$Th decay chain there are isotopes that can emit particles which produce an important background for the final detector. Six different SBD measurement setups were used at the Milano-Bicocca environmental radioactivity laboratory, which were optimized for low background measurements. Certification for surface contamination of some specific material was obtained and a continuous monitoring of the selected components was done in order to guarantee the stability in time.

### 3.2.4 Cryogenic test

A dedicated cryogenic setup mounted and operated at LNGS is used to test $TeO_2$ crystals. The tests are performed on crystals randomly chosen from each production batch and is aimed at checking the radioactive contamination level of crystals and their bolometric performance, measured in conditions similar to those planned to be used for CUORE.

### 4. Experimental results and discussion

During the $TeO_2$ crystal growth process fully described in [13], measures were taken to improve the radio-purity of reagents and intermediate products in order to reduce the bulk contamination of the grown crystals. Furthermore the equipment and consumables used for the post-growth processing of crystals were studied in order to reduce the surface contamination. The main results are discussed in the following sections.

### 4.1 Reagents and consumables

The results of the radio-purity analysis made on reagents, cleaning liquids and consumables used for $TeO_2$ crystal production are given in Table 2. Samples $H_2O$ (a) and $H_2O$ (b) refer respectively to the ultrapure water used at Kunshan Jincheng Chemical Facility for the chemical synthesis and successive drying of $TeO_2$ powder and ultrapure water used in the clean room at Jiading for the surface processing of $TeO_2$ crystals. Sample "PE bags" refers to polyethylene bags in which the abrasive $SiO_2$ powder used for the final polishing of crystal faces is delivered. They will be discussed in Section 4.4.

### 4.2 Raw material and intermediary product

The results of ICP-MS radio-purity analysis made on raw materials (metallic Tellurium) and intermediary products (different $TeO_2$ powders) are also given in Table 2. Some of these materials were also subject to HPGe measurements reported in Table 3. Besides a higher sensitivity, HPGe measurement has the advantage of measuring not only the progenitor of a radioactive chain but also the daughter concentration. In cases where secular equilibrium is broken, the ICP-MS method is able to measure only the concentration of primordial and is not able to measure the daughter content. Moreover a much higher sensitivity can be reached for the measurement of $^{40}$K concentration, which makes HPGe measurements complementary to the ICP-MS ones. This feature was crucial in the case where $^{40}$K concentration in raw Te metal was above certification limit. As table 3 shows, in the case of batch 010002, a high $^{40}$K contamination was measured in the metallic Tellurium; however, the production process was continued anyway, because the chemical synthesis of TeO$_2$ powder was expected to result in additional purification. This hypothesis was confirmed by measurements performed on the resulting TeO$_2$ (RD) powder, showing a $^{40}$K concentration within specification limits.

HPGe spectroscopy was also applied for testing the purity of powders used for the final crystal polishing. A dedicated study was done to select the polishing powder. The usual polishing powders, CeO$_2$ and Al$_2$O$_3$, were discarded because of their intrinsic radioactivity ($^{142}$Ce) or relatively high contamination in U and Th. The final choice was SiO$_2$ powder (the Admatechs product Admafine SO-E5 ), for which HPGe tests made on different production batches gave U and Th contamination levels within specification limits.

**Table 2** Typical contamination levels of raw materials, reagents, consumables and intermediary products used for the production of TeO$_2$ crystals as measured by ICP-MS technique. Acronyms RD, RC, UPD and UPC are defined in Section 3.1. The numbering 01, 02, 03 corresponds to different production batches

|  | K [ppb] | Pt [ppb] | Pb [ppb] | Bi [ppb] | Th [ppb] | U [ppb] |
|---|---|---|---|---|---|---|
| H$_2$O (a) | <10 | <0.003 | <0.1 | 0.663 | <0.001 | <0.001 |
| H$_2$O (b) | <10 | - | 0.110 | - | <0.001 | <0.001 |
| HNO$_3$ | <100 | <0.100 | 1.9 | 14 | <0.010 | <0.010 |
| HCl | <100 | 0.38 | 3.1 | 0.085 | <0.001 | <0.010 |
| NH$_4$OH | <100 | <0.003 | <0.1 | 0.991 | <0.001 | <0.001 |
| PE bags | <10 | <0.001 | <0.001 | <0.001 | <0.001 | <0.001 |
| SiO2 | <5000 | <1 | <10 | <0.5 | <0.5 | <0.5 |
| Te metal_01 | <5000 | <1 | 350 | 140 | <0.2 | <0.2 |
| TeO$_2$ RD_01 | <5000 | 6 | 32 | 25 | <0.2 | <0.2 |
| TeO$_2$ RC _01 | <5000 | 16 | 84 | 18 | ~0.3 | <0.2 |
| TeO$_2$ UPD _01 | <5000 | <15 | <200 | 51 | <0.2 | <0.2 |
| TeO$_2$ UPC _01 | <5000 | 48±12 | <100 | 67±10 | <0.2 | <0.2 |
| Te metal_02 | 41000 | <1 | 1080 | 800 | <0.2 | <0.2 |
| TeO$_2$ RD_02 | <5000 | <50 | 10 | 40 | <0.2 | <0.2 |
| TeO$_2$ RC _02 | <5000 | <50 | 27 | 37 | <0.2 | <0.2 |
| TeO$_2$ UPD _02 | <5000 | <10 | <500 | 22 | <0.2 | <0.2 |
| TeO$_2$ UPC _02 | <5000 | 70 | <500 | 22 | <0.2 | <0.2 |
| Te metal_03 | <5000 | <1 | 600 | 700 | <0.2 | <0.2 |
| TeO$_2$ RD_03 | <5000 | <10 | 55 | 10 | <0.2 | <0.2 |
| TeO$_2$ RC _03 | <5000 | 750 | <20 | 11 | <0.2 | <0.2 |
| TeO$_2$ UPD _03 | <5000 | 7 | <100 | 40 | <0.2 | <0.2 |
| TeO$_2$ UPC _03 | <5000 | 7 | <100 | 31 | <0.2 | <0.2 |

**Table 3** HPGe measurements performed at LNGS on raw materials, consumables and intermediary products used for the production of TeO$_2$ crystals. Acronyms and numbering of samples are explained in the caption of Table 2

|  | $^{232}$Th [g/g] | $^{238}$U [g/g] | $^{235}$U [g/g] | $^{40}$K [g/g] | $^{137}$Cs [mBq/kg] | $^{60}$Co [mBq/kg] |
|---|---|---|---|---|---|---|
| Te metal_01 | <6.2E-11 | <7.3E-11 | <8.7E-9 | (2.1±0.4)E-6 | 1.4±0.3 | <0.15 |
| Te metal_02 |  |  |  | (6.0±0.7)E-5 |  |  |
| Te metal_03 | <1.7E-10 | <4.0E-11 | 9.4E-9 | (2.1±0.3)E-6 | 2.5±0.4 | <0.21 |
| Te metal_04 | <1.4E-10 | <3.2E-11 | <4.8E-9 | (2.1±0.2)E-6 | <0.4 | <0.1 |
| Te metal_05 | <2.0E-10 | <3.7E-11 | <4.3E-9 | (1.7±0.2)E-6 | <0.49 | <0.21 |
| TeO$_2$ RD_01 | <1.3E-10 | <6.4E-11 | <2.7E-9 | <8.6E-8 | <0.21 | <0.28 |
| TeO$_2$ RD_02 | (7.4± 3.3)E-11 | <3.8E-11 | <5.4E-10 | (7.9± 3.2)E-8 | <0.14 | <0.16 |
| TeO$_2$ RC_05 | (4.4± 1.3)E-11 | <2.5E-11 | <7.5E-9 | (1.0± 0.3)E-7 | <0.17 | <0.19 |
| TeO$_2$ UPC_04 | (1.4± 0.2)E-11 | <1.0E-11 | <15.0E-9 | (1.4± 0.4)E-7 | <0.15 | <0.06 |
| SiO$_2$ | (1.3± 0.4)E-11 | <1.0E-10 | <0.55E-9 | (4.2± 1.0)E-7 | <0.13 | <0.10 |

**4.3 TeO$_2$ crystal growth**

The first requirement for growing radio-pure TeO$_2$ crystals is the high purity of TeO$_2$ powder used as raw material. From Tables 2 and 3, it is clear that the process of chemically synthesizing TeO$_2$ powder from Te greatly reduces the contamination from K, and to a lesser degree Pb, Bi, and Cs. The calcination process with the Pt crucible afterward was introduced to eliminate the impurities specific to wet chemical synthesis processes which evaporate at relatively low temperatures, despite the increased risk of Pt contamination. When the first crystal growth is finished, approximately 1mm is removed from the surface of the ingots, to avoid surface contamination of Pt and some other impurities. The double growth technique is used to reduce the concentration of impurities, especially radioactive elements.

The seeds used for the first batch of crystals for DBD application were previously tested for their radio-purity by a dedicated ICP-MS measurement performed on "twin slabs" cut from the crystal near the region where each seed was extracted. The seeds used for successive batches are taken only from crystals dedicated to DBD application, i.e. implicitly compliant with radio-purity specifications.

### 4.4 TeO$_2$ crystal surface processing, handling and packing

TeO$_2$ crystals to be used for CUORE experiment have very strict specifications concerning dimensions, surface quality and crystallographic orientation. Crystals have cubic shape with (50 ± 0.050) mm edges and chamfers of 0.5 mm. The average flatness of faces has to be <0.010 mm and cubic faces have to be oriented parallel to crystallographic planes ([001], [110], [1-10]) within ±1°.

Before surface processing, each crystal is cut into a cubic shape and X-ray oriented with a precision better than 1°. After that, the crystal undergoes final surface processing in a clean room in order to refine the crystal's dimensions and clean its surface, taking into account the mutual conditioning of these two targets. The crystal shape, dimensions and crystallographic orientation of faces are brought very close to their nominal values during the preliminary mechanical processing performed outside the clean room. Inside the clean room, it is difficult to greatly modify these parameters, due to strict operational constraints aimed at avoiding any radio-contamination risk. Dimensions are directly measured only once in the clean room, before etching. Afterwards, the crystal is weighed after the etching and the successive polishing of each face. The modification of dimension following each of these manufacturing steps is calculated based on the difference in weight.

At a crystal's reception in the clean room, dimensions are measured with a digital bench micrometer. Five touch points positioned respectively in the center and on the corners of each face are measured and the corresponding mean surface is calculated. Crystal dimension on one direction is computed as the distance between corresponding opposite surfaces. Planarity of one surface is calculated as the largest distance along the normal to the surface between the five touch points. Tolerances are calculated taking into consideration the largest and the shortest distance between touch points of two opposite faces along the normal to these faces. The results of planarity measurements made on the 2 hard faces (<100>) and 4 soft faces (<110>) of the first production batch of 63 crystals are given in Fig. 3 together with the dimensions of these crystals. The slightly larger mean value of crystal's dimension on the hard <100> direction is due to the difficult polishing of the corresponding (hard) faces.

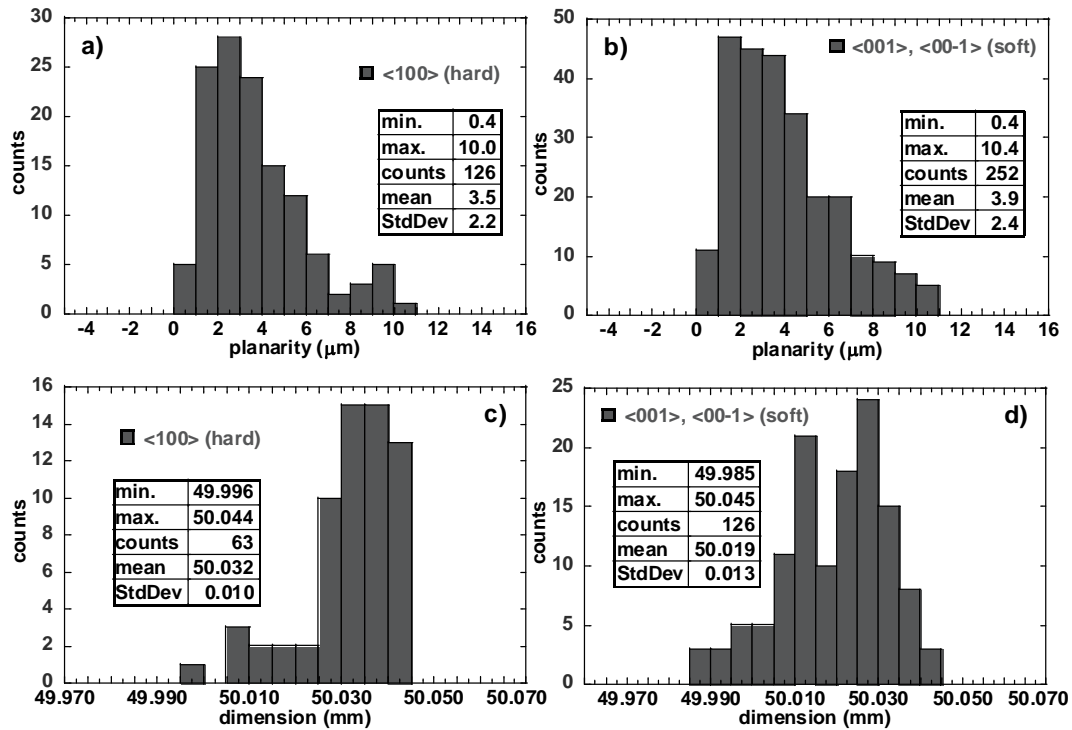

Fig. 3 Planarity and dimensions of the first production batch of 63 TeO$_2$ crystals for CUORE.

As already mentioned, the final mechanical processing also has the purpose of deep cleaning the crystal's surfaces, which may have been contaminated during the rough mechanical processing (cutting, shaping, grinding and lapping). The cleaning process is made in two steps, first by chemical etching and second by polishing. The polishing also smooths the crystal faces, possibly damaged by chemical etching. The targeted number of atomic layers to be taken away by these two procedures is on the order of $10^4$ in order to eliminate all impurity atoms that may have adsorbed on the crystal's face and further diffused in its bulk. As Fig. 4 a) shows, the mean depth of the crystal layer washed away by chemical etching is 13.5 microns and the least value of this width is 10.2 microns which makes more than $10^5$ atomic layers.

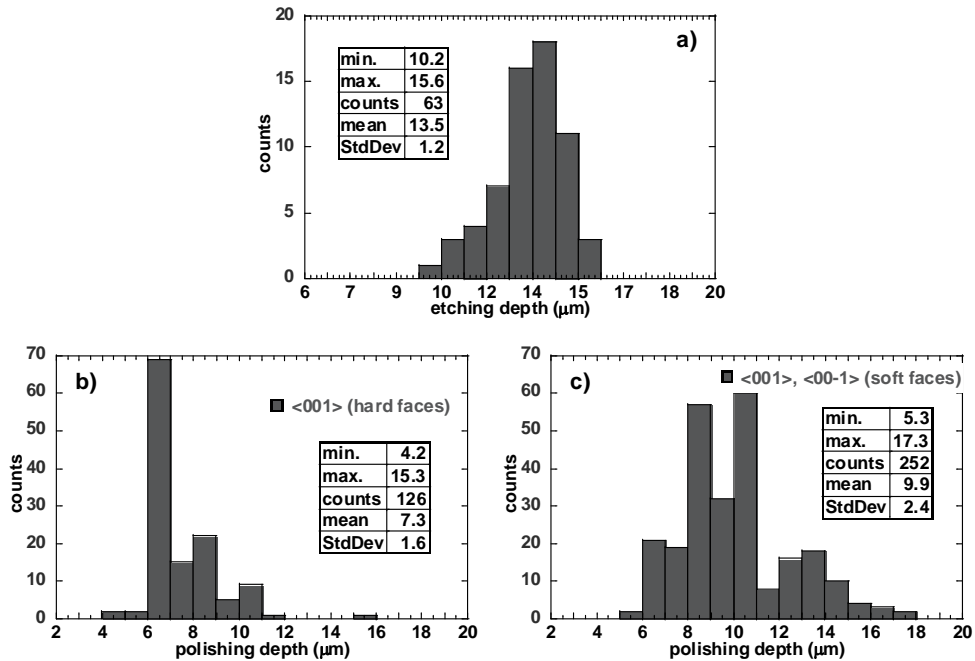

Fig. 4 Width of the layers chemically etched (a) and mechanically polished (b and c) from the surface of the $TeO_2$ crystals of the first production batch for CUORE.

Polishing is the next step in the surface cleaning procedure. As Fig. 4 b) and c) show, approximately $10^5$ atomic layers are taken away by polishing. The relatively large spread of the polishing depth is due to the fact that polishing is not only a "surface cleaning" process. Besides removing a surface layer possibly contaminated with radioactive isotopes, the polishing is also intended to bring the crystal dimensions as close as possible to the nominal values and to improve the surface quality by removing possible extended defects induced by chemical etching.

Though hard and soft faces are easily distinguishable by an expert operator, very strict rules were defined in order to avoid mistakes in the processing. Crystals are delivered to the clean room with one of the hard faces facing up, and all operations in the clean room (ultrasound washing, chemical etching and mechanical polishing) are made in a way which keeps track of the crystallographic orientation of the crystal faces.

Special care is taken in order to avoid any contamination risk during the operations in the clean room. The polishing slurry is made by adding ultrapure water directly inside the powder bags (PE bags in Table 1) and mixing the slurry in situ. Once the crystals have been etched, they come into direct contact only with the following materials: ultra clean gloves, ultra pure water, polishing slurry, polishing pads, cleaned and conditioned polyethylene sheets, a vacuum plastic bag.

The abrasive powder was selected based on dedicated research on the radio-purity and mechanical characteristics of different materials available on the market. After a preliminary study of different abrasives, the Admatechs product Admafine SO-E5 (average particle diameter 1.5μm) was selected for having the best radio-purity, even though it has moderate material removal efficiency. A special package made in a dedicated clean area and using certified polyethylene bags is guaranteed at the production site in Nagoya (Japan). The package consists of two successive bags. After removal of the outer bag the inner bag is used for the preparation of the polishing slurry by adding ultrapure water. The operation is done in the clean room just before starting the polishing operation to avoid contamination of the polishing slurry.

SBD tests were performed on consumables used in the phase of final surface processing and packaging of $TeO_2$ crystals. Polishing pads from several major producers were analyzed, and after a first phase of rough qualification, only pads produced by Lamplan and Buehler companies were selected; the other candidates were excluded for their high radio-nuclide content. In the second phase Buehler pads "Nylon 40-7068" were selected for their mechanical characteristics, which were better adapted for the use of $SiO_2$ slurry. In the last phase "Nylon 40-7068" pads belonging to different production batches were measured in order to confirm the limits of radio-contamination and to check the consistency in the production.

At the end of clean room operation, crystals are packed and barcode labeled in a triple vacuum package and stored by groups of six crystals in polyethylene vacuum boxes (Fig. 5). This special packaging system for crystal transport and storage is designed to avoid surface contamination, mainly due to radon exposure in free atmosphere. A thorough analysis of the dynamics of radon contamination was made, taking into consideration three possible processes: diffusion of radon through the packaging material, the deposition of particulates bonded with radioactive elements, and nuclear recoil implantations resulting from nuclear decay.

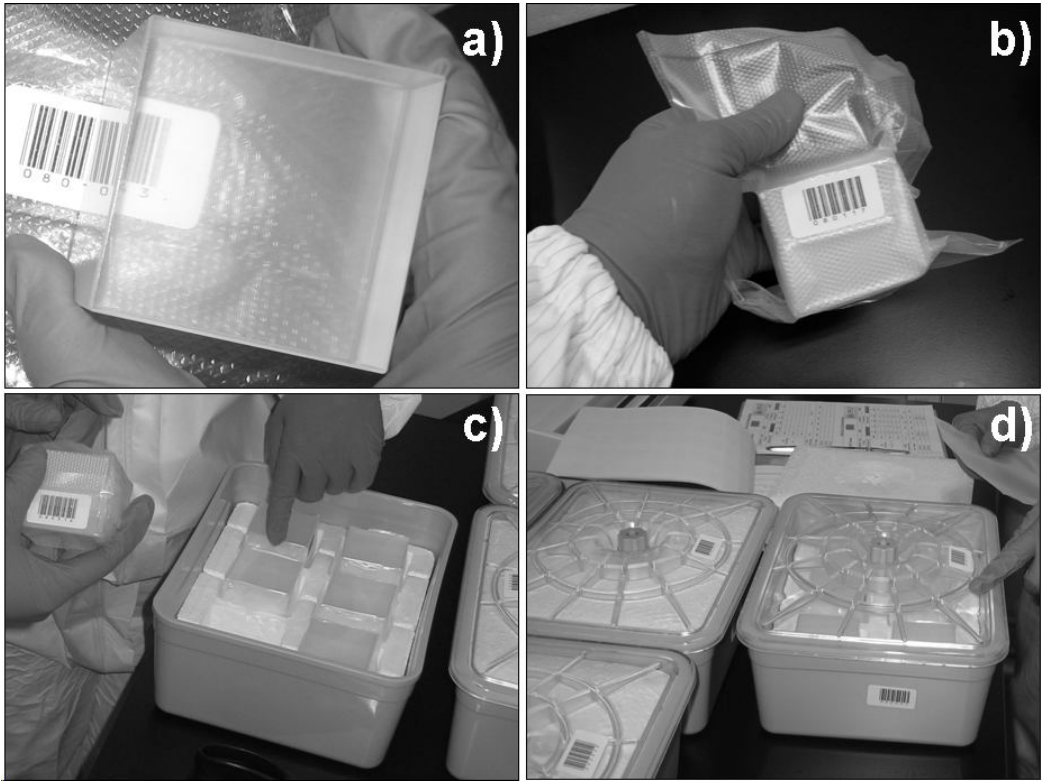

Fig. 5 TeO$_2$ crystals packaging: a) crystal ready-to-use, b) vacuum packed crystal in triple polyethylene bag, c) crystals packed in vacuum box with shock absorber "egg crate", d) vacuum boxes ready for shipment.

For long term storage, the packaged crystals are placed in a controlled environment with low-radon atmosphere (<10 Bq/m$^3$). The quality of the packaging system was checked in a radon-enriched atmosphere created inside a hermetically closed box where a uranium source was placed (radon is released in the process of uranium decay). The atmosphere was monitored with two radon detectors, one inside and one outside the box. The surface contamination was studied inside this box on four identical slabs of copper: one unpackaged, and one each packed in one, two, or three successive polyethylene bags. A fifth identical slab was kept in safe (dry, dust free) conditions outside the box in ordinary atmosphere as a "blank". Copper samples were used instead of TeO$_2$ because of their higher chemical activity. The samples were kept inside the box for approximately 20 days. The Radon concentration monitored during the exposure time was 350 kBq/m$^3$, which is five orders of magnitude higher than expected in real storage conditions. The analysis of the slabs after the exposure to this radon enriched atmosphere clearly showed that the packaging with 3 polyethylene bags is safe, i.e. the radon does not diffuse through the polyethylene and the packaging used is sufficient to prevent the contamination of the crystals.

### 4.5 Cryogenic tests

The cryogenic tests referred in this work were performed on 4 crystals randomly chosen from a set of 67 belonging to two different production batches. The crystals showed excellent bolometric characteristics including very good energy resolution (FWHM on the order of 5 keV in the spectral region of interest for CUORE). Preliminary analysis also showed that for an acquisition time corresponding to 52.6 days live time, the bulk contamination was measured to be $1.8 \times 10^{-14}$ g/g for $^{238}$U and $5.5 \cdot 10^{-14}$ g/g for $^{232}$Th at 90% C.L. and assuming secular equilibrium. These values are well below the concentration limits requested for TeO$_2$ crystals to be used in CUORE experiment (Table 1). The energy spectra of coincident events (simultaneous signals from neighboring crystals in the measurement set-up) were used to estimate the surface contamination of the crystals. The value measured (90% C.L.) was $7.9 \cdot 10^{-9}$ Bq/cm$^2$ for $^{238}$U and $1.1 \cdot 10^{-8}$ Bq/cm$^2$ for $^{232}$Th.

### 5. Conclusion

We have successfully grown high purity TeO$_2$ crystals and used them to produce TeO$_2$ crystal bolometers compatible with the challenging requirements of the CUORE DBD experiment.

Dedicated crystal growth and surface processing technologies were used to obtain crystals with excellent crystallographic characteristics, reflected by their very good energy resolution as bolometers (FWHM of the order of 5 keV in the spectral region of interest for CUORE experiment).

Surface treatments made in clean room conditions using specially selected reagents for chemical etching and selected consumables for the final polishing resulted in crystal surface contamination of $7.9 \cdot 10^{-9}$ Bq/cm$^2$ for $^{238}$U and $1.1 \cdot 10^{-8}$ Bq/cm$^2$ for $^{232}$Th.

High sensitivity measurements of radio-isotope concentrations in raw materials, reactants, consumables, ancillaries and intermediary products and the application of very strict production and certification protocols led to a very low radio-contamination level measured in ready-to-use TeO$_2$ crystals. The bulk contamination measured in crystals randomly chosen from different production batches was $1.8 \cdot 10^{-14}$ g/g for $^{238}$U and $5.5 \cdot 10^{-14}$ g/g for $^{232}$Th, well below the concentration limits requested for TeO$_2$ crystals to be used in CUORE experiment.

**Role of the funding sources**

This work was supported by the US Department of Energy under contract numbers DE-AC52-07NA27344 at LLNL and DE-AC02-05CH11231 at LBNL, and by the INFN of Italy in the frame of CUORE Collaboration.